\documentclass[twoside, 11pt]{article}
\usepackage[english]{babel}
\usepackage{epsfig}
\usepackage{psfig}
\usepackage{amssymb,amsmath,amsfonts,soul,amsthm,enumerate}
\usepackage{graphicx}
\usepackage[cp1251]{inputenc}
\usepackage[T2A]{fontenc}
\usepackage{mathrsfs}
\usepackage{booktabs}
\usepackage{float} 
\usepackage{algorithm}
\usepackage{algorithmic}
\usepackage{listings}%

\usepackage{orcidlink}

\def\@evenfoot{}
\def\@oddfoot{}
\begin{document}
\def\@evenhead{\vbox{\hbox to \textwidth{\thepage\leftmark}\strut\newline\hrule}}

\def\@oddhead{\raisebox{0pt}[\headheight][0pt]{%
\vbox{\hbox to \textwidth{\rightmark\thepage\strut}\hrule}}}


\newpage
\normalsize
\def\bibname{\vspace*{-30mm}{\centerline{\normalsize References}}}
\thispagestyle{empty}
\begin{flushleft}
{\small {\bf EURASIAN  JOURNAL OF MATHEMATICAL \\ AND COMPUTER APPLICATIONS}\\
 ISSN 2306--6172\\
Volume  13, Issue 3  (2025) 4 -- 21}
\end{flushleft}
\vskip 5 mm

\centerline{\bf PLATOS: A POWER AND LATENCY AWARE TASK}
\centerline{\bf ORIENTED SCHEDULING STRATEGY FOR HEALTHCARE IOT}
\centerline{\bf IN FOG COMPUTING}

\vskip 0.2cm
\centerline{\bf Ala'anzy M.A. \orcidlink{0000-0002-0005-7037}\footnotemark[1] , Ahmad Z. \orcidlink{0000-0002-6278-9516} , Mukash Zh. \orcidlink{0009-0004-7713-3925} }
\footnotetext[1]{Corresponding Author.}
\vskip 0.2cm
\noindent{\bf Abstract}
{\small Healthcare Internet of Things (HIoT) technology is revolutionising the healthcare industry by enabling real-time data collection and analysis for personalised patient care. However, the rapid expansion of HIoT technology introduces challenges such as increased latency and higher energy consumption in fog computing environments, particularly when managing battery-operated devices. To address these issues, this work proposes a novel scheduling strategy that optimises both power consumption and latency through task-oriented scheduling for HIoT tasks. The proposed strategy, named PLATOS (Power and Latency Aware Task Oriented Scheduling), is implemented in four sequential phases. In the first phase, HIoT tasks are categorised into three groups: priority-oriented, storage-oriented, and computational-oriented. The second phase focuses on latency optimisation by identifying the fog computing resources that yield the lowest execution delay for each task category. In the third phase, power optimisation is achieved by selecting the resources that minimise energy consumption. Finally, in the decision-making phase, high-performance fog resources are allocated to high-priority tasks while the remaining tasks are scheduled based on a mapped list derived from the latency and power optimisation phases. Simulation experiments conducted in iFogSim2 demonstrate that PLATOS reduces energy consumption by 18.72\% and latency by 8.65\% when compared to the state-of-the-art. These improvements enhance the efficiency and responsiveness of HIoT systems and contribute to more effective patient care and proactive healthcare service delivery.
\vskip 0.2cm
\noindent {\bf Keywords:} Cloud computing, fog computing, healthcare, internet of things, energy consumption, latency.
\vskip 0.2cm
\noindent {\bf AMS Mathematics Subject Classification:} 90B35, 90C05, 68M14, 68U20.}
\vskip 0.2cm
\noindent {\bf DOI:}  10.32523/2306-6172-2025-13-1-4-21
\vskip 0.3cm
\setcounter{figure}{0}

\section{Introduction}\label{Sec: Introduction}\vspace{-2mm}

Fog computing extends cloud computing capabilities to the network edge, enabling computational resources to be positioned closer to data sources. This architecture addresses several limitations of traditional cloud computing, including data privacy concerns, bandwidth constraints, and latency issues \cite{srirama2024decade}. Over time, fog computing has been integrated with the Internet of Things (IoT) across various domains, including atmospheric attenuation analysis for rain data assessment \cite{li2023atmospheric, li2022attenuation}, smart cities \cite{ala2024real}, smart building solutions for individuals with disabilities \cite{safi2024design}, security applications \cite{algarni2024edge}, healthcare, and more.

In the healthcare sector, the Healthcare Internet of Things (HIoT) refers to the integration of IoT-enabled medical devices, wearables, sensors, and other connected technologies that facilitate real-time health data collection and transmission \cite{panda2024medical, sivakumar2024addressing}. This integration enhances patient monitoring, enables early diagnosis, and supports data-driven decision-making for improved healthcare delivery.

Fog computing significantly enhances the HIoT ecosystem by processing and analysing data near the data source. This edge processing reduces latency, accelerates data processing, and minimises the volume of data transmitted to the cloud \cite{wen2024exploitation}. In critical healthcare applications like remote patient monitoring, telemedicine, and emergency response, timely data processing is crucial. Fog computing ensures that vital health data is processed promptly, thereby improving patient outcomes and reducing the risk of adverse events. Thus, fog computing has the potential to transform healthcare IoT by enabling more efficient and effective data processing and analysis, leading to enhanced patient care and a more responsive healthcare system \cite{rani2024integration,ala2024dynamic}.

Task allocation in an IoT environment involves distributing tasks and resources among various IoT devices in a network. The heterogeneity of IoT devices, in terms of capabilities, makes task allocation challenging. It is essential to optimise this process to assign tasks to devices that can execute them efficiently while minimising resource usage \cite{al2023energy}. Various approaches to task allocation exist. Centralised allocation involves a central entity managing task distribution based on device capabilities and availability, which is straightforward but can suffer from scalability and latency issues \cite{tarek2020new}. In contrast, decentralised allocation allows each IoT device to independently decide which tasks to accept and how to allocate its resources, offering greater flexibility and scalability but requiring more sophisticated algorithms for efficient resource use \cite{mijuskovic2021resource}. Hybrid approaches that combine centralised and decentralised methods are also employed, depending on the specific requirements of the IoT environment \cite{patsias2023task}. Effective task allocation is crucial for IoT network management and optimisation, and the appropriate approach should be selected based on the unique characteristics of the IoT setup \cite{dayong2024comprehensive}.

Power consumption is a critical factor in the task allocation process within an IoT environment. Many IoT devices are battery-powered and have limited energy resources, necessitating careful power management to ensure device longevity and optimal performance. Efficient task allocation minimises power consumption by assigning tasks to devices capable of performing them with minimal energy use \cite{aldin2023comprehensive}. For instance, tasks requiring high processing power or data transmission might be allocated to devices with more powerful processors or superior communication capabilities, while less demanding tasks could be assigned to less powerful devices. Additionally, energy-efficient algorithms and techniques, such as data compression, aggregation, and dynamic voltage and frequency scaling, are employed to further reduce power consumption during task execution \cite{ala2022mapping,alanzy2018range,ala2020optimising}. By managing power consumption through efficient task allocation and energy-saving techniques, IoT devices can perform their tasks optimally, ensuring extended battery life and enhancing the overall efficiency and effectiveness of the IoT network \cite{alsharif2024comprehensive}.

Energy-efficient and task-oriented resource allocation for HIoT data in fog environments presents several challenges that must be addressed to achieve optimal resource utilisation and task performance \cite{mijuskovic2021resource}. One of the primary challenges is the energy constraints of many IoT devices, which limit their ability to perform complex tasks or transmit large volumes of data. Developing energy-efficient resource allocation strategies is essential to ensure these devices can operate for extended periods without depleting their power reserves.

In healthcare IoT applications, low latency is crucial to ensure that patient data is transmitted and processed swiftly. Allocating fog computing resources to minimise latency is challenging due to the distributed nature of fog environments. Furthermore, healthcare IoT data is often sensitive and requires protection against unauthorised access or tampering. Resource allocation strategies must consider these security requirements, which may involve additional processing and encryption measures \cite{rani2024integration}.

The heterogeneity of healthcare IoT devices, which vary in processing capabilities, memory, and network connectivity, adds another layer of complexity to resource allocation. Strategies must accommodate this diversity and allocate resources effectively. Scalability is also a significant challenge, as healthcare IoT applications generate vast amounts of data that need to be processed efficiently. Resource allocation strategies must be scalable to handle large data volumes and support future growth in the number of IoT devices and data \cite{raghavendar2023robust}.

Addressing these challenges requires treating energy-efficient and task-oriented resource allocation for healthcare IoT data in a fog environment as a complex optimisation problem. This problem necessitates careful consideration of energy constraints, latency requirements, security concerns, device heterogeneity, and scalability \cite{diyan2022intelligent}. By employing intelligent resource allocation algorithms, it is possible to optimise resource utilisation and ensure the efficient processing of healthcare IoT data in fog computing environments.

This research proposes a PLATOS strategy for HIoT tasks. The main contributions are outlined below:
\begin{enumerate}\itemsep=-2pt
    \item The research presents a framework that implements power and latency optimisation through task-oriented scheduling for HIoT tasks in fog computing. Power and latency are optimised through the following phases:
    \begin{enumerate}\itemsep=-2pt
        \item \textbf{Task-orientation phase:} This phase categories HIoT tasks into three categories: priority-oriented tasks, storage-oriented tasks, and computational-oriented tasks.
        \item \textbf{Latency-optimisation phase:} This phase identifies fog resources for each category of tasks in ascending order of minimum latency during execution.
        \item \textbf{Power-optimisation phase:} This phase identifies fog computing resources for each category of tasks in ascending order of minimum power consumption.
        \item \textbf{Decision-making phase:} This phase schedules high-performance fog resources for high-priority HIoT tasks, while the remaining HIoT tasks are scheduled using a mapped list of resources from the latency and power-optimisation phases.
    \end{enumerate}
    \item The proposed strategy is evaluated through a simulation environment, with performance evaluation parameters including energy consumption and latency. The simulation results are compared with existing state-of-the-art strategies.
\end{enumerate}
	
The paper is organised as follows: Sec.~\ref{Sec: Related Works} discusses related work that is relevant to the proposed research. Sec.~\ref{Sec: System Design and Model} provides details on the system design and model. The evaluation methods are described in Sec.~\ref{Sec: Evaluation Methods}. Sec.~\ref{Sec: Experiments, Results, and Discussion} provides the experimental setup, results, and discussion. Sec.~\ref{Sec: Conclusion} concludes the article, highlighting limitations and suggesting directions for future work.\vspace{-2mm}

\section{Related Works}\label{Sec: Related Works}\vspace{-2mm}

The related work is studied and analysed with respect of fog computing, HIoT, energy-efficient, and task-oriented scheduling algorithms. The IoT has emerged as a key technology that has the potential to revolutionise various industries, including healthcare. Remote patient monitoring through IoT devices can help healthcare providers to monitor the health of patients remotely and in real-time. However, the latency caused by transferring data from sensors to the cloud and back, be a significant challenge in remote health monitoring applications. To overcome this challenge, fog computing can be used as an intermediate layer between sensors and cloud computing, which collects and processes data more efficiently and reduces the amount of data transferred between sensors and the cloud. Wireless sensor networks (WSNs) are commonly used in healthcare applications to collect and transmit data from IoT devices to fog and cloud computing systems. However, these networks often send a large number of tasks simultaneously, which results in task delays and decreased system performance. Therefore, an appropriate task-scheduling algorithm is needed to prioritise tasks and ensure that high-priority tasks are processed quickly, regardless of their length. In \cite{aladwani2019scheduling}, the authors propose a new method called Tasks Classification and Virtual Machines Categorisation (TCVC) based on task importance to improve the performance of static task scheduling algorithms. The proposed method classifies tasks based on their importance into three categories: high importance, medium importance, and low importance, depending on the patient's health status. The proposed research aims to improve the performance of task scheduling algorithms in healthcare applications using IoT and fog computing. By prioritising tasks based on their importance, the proposed method helps to ensure that high-priority tasks are processed quickly and efficiently, even if they are long tasks. The use of fog computing as an intermediate layer also helps to reduce latency and improve system efficiency.

The research work in \cite{jamil2020job} presents a new scheduler for Fog computing that optimises network usage and delay specifically for the Internet of Everything (IoE) environment. Fog computing offers storage, processing, analytical, and networking services at the network edge, resolving the latency and bandwidth issues that Cloud computing faces. Nevertheless, Fog devices at the network edge have limited resources, making job scheduling and resource allocation challenging tasks. A well-designed job scheduling algorithm minimises energy consumption and response time for application requests. The proposed scheduling algorithm was tested using iFogSim and demonstrated improved network usage and delay compared to existing approaches. 

In \cite{attiya2022intelligent}, the authors highlight the requirements in cloud computing service usage due to the expansion of IoT-based applications and the need for intelligent scheduling methods to optimise the scheduling of IoT application tasks on computing resources. The authors suggest a novel algorithm, CHMPAD, which integrates the chimp optimisation algorithm (ChOA), marine predators algorithm, and disruption operator to prevent being trapped in local optima and enhance the exploitation capability of the basic ChOA. The simulation results reveal that CHMPAD significantly enhances the makespan time and throughput performance of fog computing when compared to other scheduling algorithms. The proposed algorithm is validated using synthetic and real workloads gathered from the Parallel Workload Archive.

In \cite{azizi2022deadline}, the authors address the challenge of efficient deployment of fog computing resources for executing heterogeneous and delay-sensitive IoT tasks. The authors propose a mathematical model for task scheduling that considers the minimisation of energy consumption, deadline violation time, and Quality of Service (QoS) requirements of IoT tasks. Two semi-greedy-based algorithms, PSG and PSG-M, are proposed to map IoT tasks to fog nodes efficiently. The performance evaluation shows that the proposed algorithms outperform existing algorithms in terms of meeting the deadline requirement, reducing deadline violation time, optimising energy consumption, and makespan of the system. This paper provides a valuable contribution to the efficient deployment of fog computing resources for IoT applications. 

The proposed Energy-Efficient Internet of Medical Things to Fog Interoperability of Task Scheduling (EEIoMT) framework by \cite{azizi2022deadline}, appears to be a promising solution for real-time healthcare applications that utilise fog computing. The framework addresses the critical need for efficient task scheduling that minimises response time, latency, and energy consumption while ensuring that priority-based tasks are executed within their deadline. The architecture described in the study utilises ECG sensors to monitor heart health at home and sends the sensed data to the fog scheduler for analysis. The scheduler selects the appropriate fog node based on a weighted formula that considers the expected energy consumption and latency of executing each task. Simulation results suggest that the proposed framework outperforms existing models such as CHTM, LBS, and FNPA in terms of reducing energy usage, latency, and network utilisation. The study provides valuable insights into the potential benefits of fog computing in healthcare and the importance of developing efficient task scheduling algorithms to optimise its use. However, further research and testing are needed to validate the proposed framework's effectiveness in real-world settings and its potential impact on improving patient outcomes.

Fog computing is a promising solution to address the challenges of bandwidth, network latency, and energy consumption faced by cloud computing. Healthcare IoT devices generate massive amounts of data that need to be efficiently managed with minimal latency, energy consumption, and cost. Failures in tasks or nodes increase latency, energy consumption, and cost, which can have severe consequences for patients. To address these challenges, a ``Fault Tolerant Data Management'' (FTDM) scheme has been proposed for healthcare IoT in fog computing \cite{saeed2021fault}. FTDM efficiently organises and manages healthcare IoT data through well-defined components and steps. The scheme includes a mechanism that works in two-way and manages task and node failures. Simulation results using iFogSim show significant improvements compared to the existing Greedy Knapsack Scheduling (GKS) strategy. The FTDM scheme is particularly valuable in circumstances where patients need to be treated remotely, such as during outbreaks of infectious diseases like COVID-19. Overall, the proposed strategy is a cost-efficient, energy-aware, and fault-tolerant approach for managing healthcare IoT data in fog computing, which improves system performance and saves patients' lives by minimising latency and providing fault tolerance.

Recent studies have highlighted the limitations of offloading IoT tasks to the cloud, especially under conditions such as resource contention and varying provisioning levels. \cite{al2023energy} examined traditional resource scheduling algorithms, noting their focus on cost minimisation and resource optimisation, while often overlooking energy consumption as a crucial optimisation factor. They introduced a novel cooperative energy-aware resource allocation and scheduling strategy, leveraging the Technique for Order of Preference by Similarity to Ideal Solution (TOPSIS) method. Their TOPSIS-based Resource Allocation (TOPREAL) approach significantly enhances energy savings and execution time efficiency, outperforming existing algorithms. However, their methodology did not address latency considerations.

A study by \cite{saif2023multi} introduces the Multi-Objectives Grey Wolf Optimiser (MGWO) algorithm, which aims to minimise QoS objectives such as delay and energy consumption by efficiently distributing tasks via a fog broker. This algorithm shows promise in reducing both transmission delays when tasks are sent to the fog and energy consumption when offloading tasks to the cloud. The simulation results demonstrate the MGWO algorithm's superiority over state-of-the-art algorithms in enhancing performance metrics. However, while the MGWO algorithm offers improvements in delay and energy consumption, it may not comprehensively address the dynamic nature of IoT environments where task priorities can shift rapidly. Moreover, the focus on minimising delay and energy consumption could overlook other crucial aspects such as resource allocation efficiency and real-time adaptability. In comparison, our scheduling (PLATOS) strategy offers a more holistic approach by categorising tasks and optimising resources based on specific task requirements.

The integration of cloud and fog computing in the vehicular ad hoc network (VANET) and the Internet of Vehicles (IoV) is addressed in a study by \cite{ehtisham2024internet}, which introduces a fuzzy logic-based task scheduling system in VANET. This system aims to reduce latency and improve response times when offloading tasks in IoV. It manages the transfer of workloads to the fog computing layer effectively, considering the limited processing power, bandwidth, and high-speed mobility of vehicles. The system demonstrates superiority over existing algorithms, notably reducing average latency. However, while the fuzzy logic-based approach offers significant improvements, it may not fully address the complex and dynamic nature of vehicular networks, where task priorities and resource availability can fluctuate rapidly.

A study by \cite{khan2024energy} addresses the challenges of n-tier fog computing frameworks for IoT applications, focusing on the immediate execution of sensor-generated data with minimal delay and energy consumption. The study highlights the issues of fog device failure and its impact on system performance. To mitigate these issues, the authors propose an energy-efficient task scheduling algorithm based on reactive fault tolerance. Developed using modified particle swarm optimisation, this algorithm reschedules tasks to other executable fog nodes upon device failure. The proposed technique aims to reduce energy consumption, latency, and network bandwidth utilisation while increasing system reliability and success rate. While this approach offers a robust solution for enhancing fog computing performance, it may face scalability challenges as the number of fog devices or IoT applications increases. Additionally, relying solely on reactive fault tolerance may not be sufficient in dynamic IoT environments, where proactive measures could provide further benefits.

A study by \cite{thakur2024deadline} addresses the challenge of efficiently assigning tasks in fog computing to minimise makespan and energy consumption while maximising the number of tasks meeting their deadlines. The study proposes an enhanced semi-greedy algorithm, integrating fuzzy logic to improve decision-making under varying conditions and uncertainties in the fog environment. Simulation experiments demonstrate that this algorithm surpasses the Priority-aware Semi-Greedy (PSG) and PSG-MultiStart (PSG-M) algorithms in reducing makespan and energy consumption and increasing deadline adherence. This flexible approach and nuanced task scheduling offer a significant advantage in complex fog computing scenarios. While the integration of fuzzy logic enhances task scheduling efficiency, the approach may need further testing across diverse real-world conditions to validate its robustness and adaptability in various fog environments.

In \cite{premalatha2024optimal}, authors proposed an Optimal Energy-efficient Resource Allocation (OEeRA) algorithm, building on the Minimal Cost Resource Allocation (MCRA) and Fault Identification and Rectification (FIR) algorithms. This approach ensures effective task offloading in IoT-FoG computing networks, assigning at least one fog node (FN) and resource block (RB) per device. Faulty RBs are replaced using stored backups, improving processing and response time and increasing fault detection accuracy. The OEeRA algorithm demonstrates significant energy efficiency gains across varying configurations of FNs, RBs, and IoT devices. However, the OEeRA algorithm shows promise in enhancing energy efficiency and fault tolerance, further research is needed to explore its scalability and adaptability in real-world, dynamically changing IoT environments.

The use of IoT devices in healthcare has the potential to revolutionise the industry, but remote patient monitoring faces challenges due to latency caused by data transfer. Fog computing is used as an intermediate layer to overcome this challenge, and task scheduling algorithms are needed to prioritise tasks and ensure high-priority tasks are processed quickly. Several studies propose new algorithms to improve the performance of static task scheduling in healthcare applications using IoT and fog computing. These studies evaluate proposed methods using simulation and show improved delay, network usage, energy consumption, and system efficiency. The proposed frameworks and algorithms provide valuable insights into the potential benefits of fog computing in healthcare but still need improvement in resource scheduling with important parameters of healthcare such as task orientation, energy efficiency and latency.\vspace{-2mm}

    \section{System Design and Model}\label{Sec: System Design and Model}\vspace{-2mm}
\subsection{Cloud-Fog System Architecture}\vspace{-2mm}

\begin{figure}[h!]
    \centering
    \includegraphics[width=90mm]{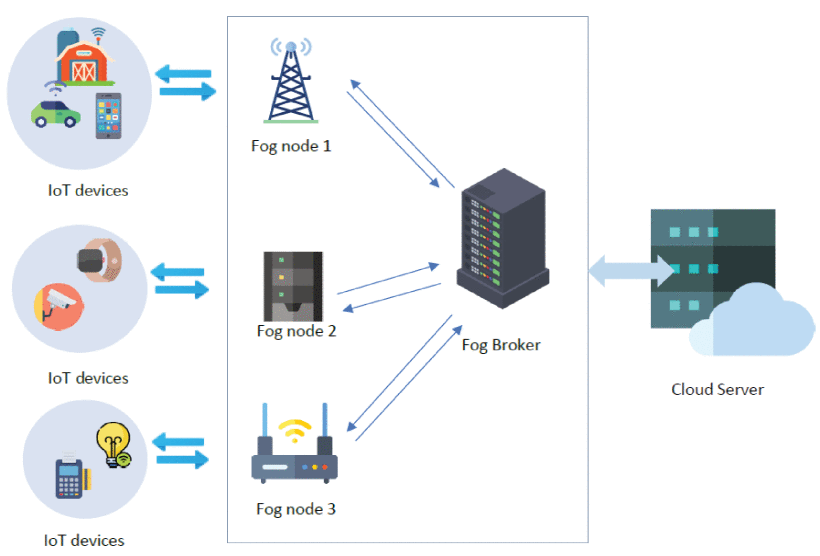}\vspace{-4mm}
    \caption{Cloud-Fog system architecture \cite{saif2023multi}.} \label{Fig: Cloud-FOG System Architecture}\vspace{-4mm}
\end{figure}

The Cloud-Fog System Architecture comprises terminal devices, $N$ fog nodes, and $C$ cloud servers, as illustrated in Fig.~\ref{Fig: Cloud-FOG System Architecture}. 
Terminal devices, connected via a wireless channel, send sensor-generated data directly to fog nodes. These nodes, in turn, forward the data to a nearby fog broker for analysis, estimation, and scheduling of end-user requests. Based on the task characteristics, the broker decides whether the fog or cloud devices are more suitable for task execution. Proximity to the fog nodes minimises time consumption at the broker. To ensure optimal task scheduling that meets both transmission delay and energy consumption criteria, the proposed algorithm is employed at the broker. The traffic model, considering the varied power and capacity of resources, views terminal devices as an $M/M/1$ queue, fog nodes as an $M/M/C$ queue, and cloud servers as an $M/M/\infty$ queue.\vspace{-2mm}

\subsection{Mathematical Model for Task Scheduling Algorithm}\vspace{-2mm}

The proposed framework is structured into three processing layers: microcontrollers, fog nodes, and the cloud. In the fog layer, each node hosts fog servers, which function as micro data centres or VMs. These servers vary in capacity across different layers. Specifically, fog nodes possess significantly less computational power, storage, and server capacity than the fog cloud, but they offer more than the microcontroller. However, when considering delays, response time, and proximity to end users, computing nodes that are closer to the data source exhibit lower latency and faster response times. Consequently, it is essential to execute critical and time-sensitive tasks on the appropriate computational node to minimise delays and ensure deadlines are met.

In fog computing, one of the primary challenges in task scheduling is to efficiently allocate IoT tasks to the most suitable fog nodes from the available options, aiming to optimise QoS. This study focuses on latency and energy consumption as key QoS parameters.

Assume there are $n$ tasks $T$ that need to be delivered to the fog scheduler, which can be expressed as follows:
\begin{equation} \label{Equ: Tasks}
    T = \{t_1, t_2, \dots, t_n\},
\end{equation}
where each task $t_i$ is characterised by a set of attributes $t_i ={TS_i, TL_i, \textit{type}_i, dt_i}$. Here, $TS_i$ represents the task size (in bits), $TL_i$ is the task length (in Millions of Instructions, MI), $\textit{type}_i$ indicates whether the task is normal, moderate, or critical, and $dt_i$ is the deadline by which the task must be completed.

Now, consider that the fog computing system consists of $m$ fog nodes $F$, which can be expressed as:
\begin{equation}\label{Equ: Fog nodes}
F = \{f_1, f_2, \dots, f_m\},
\end{equation}
where each fog node $f_j$ is described by a set of attributes $f_j = \{S_j, CC_j, E_j\}$. In this context, $S_j$ denotes the storage capacity, $CC_j$ refers to the computing capacity (in Millions of Instructions Per Second, MIPS), and $E_j$ is the total energy or battery capacity of the fog node $f_j$.

The task scheduling challenge involves allocating $n$ tasks to $m$ fog nodes in such a way that the QoS parameters are optimised, using the notations defined above. Let $X_{ij}$ represent the assignment of task $t_i$ to fog node $f_j$, while $X_{icloud}$ denotes the assignment of task $t_i$ to the cloud.

We can analytically evaluate execution time, transmission time, response time, and energy consumption to determine the optimal distribution of the tasks to the appropriate fog node. Following this analysis, the most suitable node is selected, and the task is assigned accordingly.

\textbf{Execution Time:} The execution time of processing task $t_i$ on fog node $f_j$ or the cloud is calculated using Eqs.~\eqref{Equ: ExecutionTimeFog} and \eqref{Equ: ExecutionTimeCloud} respectively:

\begin{equation}\label{Equ: ExecutionTimeFog}
Et(X_{ij}) = \frac{TL_i}{CC_j},
\end{equation}
\begin{equation}\label{Equ: ExecutionTimeCloud}
Et(X_{i_{\text{cloud}}}) = \frac{TL_i}{CC_{\text{cloud}}},
\end{equation}
where $Et(X_{ij})$ and $Et(X_{i_{\text{cloud}}})$ represent the execution time of the task on the fog node and cloud, respectively. $CC_j$ and $CC_{\text{cloud}}$ denote the computation capacity of the fog node and the cloud, respectively.

\textbf{Transmission Time:} The transmission time of task $t_i$ from the microcontroller to the fog scheduler $T_rt(FS)$ is calculated by dividing the task size $TS_i$ by the transmission rate (bandwidth) $BW$:
\begin{equation}\label{Equ: TransmissionTimeMicro2Fog}
T_rt(FS) = \frac{TS_i}{BW}
\end{equation}

The transmission time of task $t_i$ from the fog scheduler to the fog node $T_rt(X_{ij})$ or to the cloud $T_rt(X_{i_{\text{cloud}}})$ is calculated as in Eqs.~\eqref{Equ: TransmissionTimeMicro2Fog} and \eqref{Equ: TransmissionTimeSch2Cloud}:
\begin{equation}\label{Equ: TransmissionTimeSch2Fog}
T_rt(X_{ij}) = \frac{TS_i^{\text{send}} + TS_i^{\text{response}}}{BW},
\end{equation}
\begin{equation}\label{Equ: TransmissionTimeSch2Cloud}
T_rt(X_{i_{\text{cloud}}}) = \frac{TS_i^{\text{send}} + TS_i^{\text{response}}}{BW},
\end{equation}

The total transmission time of tasks from the microcontroller to the appropriate fog node $f_j$ or the cloud can be calculated by combining Eqs.~\eqref{Equ: TransmissionTimeMicro2Fog} and \eqref{Equ: TransmissionTimeSch2Fog} or \eqref{Equ: TransmissionTimeMicro2Fog} and \eqref{Equ: TransmissionTimeSch2Cloud}, respectively, as per following Eqs.~\eqref{Equ: mergeMicro and Fog} and \eqref{Equ: MergeFog&Cloud}:
\begin{equation}\label{Equ: mergeMicro and Fog}
T_rttotal(X_{ij}) = T_rt(FS) + T_rt(X_{ij}),
\end{equation}
\begin{equation}\label{Equ: MergeFog&Cloud}
T_rttotal(X_{i_{\text{cloud}}}) = T_rt(FS) + T_rt(X_{i_{\text{cloud}}}).
\end{equation}

\textbf{Formulation for Task Scheduling Problem:} The objective of task scheduling is to allocate IoT tasks to the resources of fog nodes or the cloud in the most efficient manner, with the aim of minimising latency and energy consumption. This problem can be formulated using Integer Linear Programming (ILP) \cite{pochet2006production} to represent the assignment of IoT tasks to appropriate fog nodes, as follows:
\begin{equation}\label{Equ: task is not assigned to more than one fog}
\sum_{j=1}^{m} X_{i,j} = 1 \quad \forall i \in \{1, \dots, n\};
\end{equation}
\begin{equation}\label{Equ: computation intensity}
\sum_{i=1}^{n} X_{i,j} \times C_i \leq CC_j \quad \forall j \in \{1, \dots, m\};
\end{equation}
\begin{equation}\label{Equ: task size necessary}
\sum_{i=1}^{n} X_{i,j} \times TS_i \leq S_j \quad \forall j \in \{1, \dots, m\};
\end{equation}
\begin{equation}\label{Equ: fog energy consumption}
\sum_{i=1}^{n} X_{i,j} \times E_p(X_{ij}) \leq E_j \quad \forall j \in \{1, \dots, m\};
\end{equation}
\begin{equation}\label{Equ: total time required by fog node}
RT(X_{ij}) \leq dt_i \quad \forall i \in \{1, \dots, n\}, \quad \forall j \in \{1, \dots, m\};
\end{equation}
\begin{equation}\label{Equ: defines our binary decision variables}
X_{ij} \in \{0, 1\} \quad \forall i \in \{1, \dots, n\}, \quad \forall j \in \{1, \dots, m\};
\end{equation}
Eq.~\eqref{Equ: task is not assigned to more than one fog} ensures that a task is not assigned to more than one fog node simultaneously. Eq.~\eqref{Equ: computation intensity} indicates that the computational demand required to execute the assigned tasks must not exceed the computing capacity of the fog node. Eq.~\eqref{Equ: task size necessary} ensures that the cumulative task size of the tasks assigned to a fog node does not surpass its storage capacity. Eq.~\eqref{Equ: fog energy consumption} guarantees that the energy consumption required to complete the assigned tasks does not exceed the fog node's available battery capacity. Eq.~\eqref{Equ: total time required by fog node} ensures that the total response time for fog node $f_j$ to complete task $t_i$ does not exceed the task's deadline. Finally, Eq.~\eqref{Equ: defines our binary decision variables} defines the binary decision variables, where $X_{ij}$ is 1 if fog node $f_j$ is selected to perform task $t_i$, and 0 otherwise.\vspace{-2mm}

\subsection{The PLATOS Algorithm}\vspace{-2mm}

This work introduces a novel strategy called power and latency-aware task-oriented scheduling for handling HIoT tasks. The PLATOS strategy is structured into four distinct stages to ensure efficient task management and resource optimisation.

In the first stage, known as the task-orientation phase, tasks associated with HIoT are categorised into three distinct groups: priority-oriented, storage-oriented, and computationally-oriented tasks. This classification facilitates targeted scheduling and resource allocation for each type of task, ensuring that tasks are processed in alignment with their specific requirements. Priority-oriented tasks are those with stringent deadlines or significant impact on overall system performance and are therefore scheduled first. Storage-oriented tasks are those that demand significant memory resources, whereas computationally-oriented tasks require intensive processing power. By segregating tasks in this manner, the system can allocate resources more efficiently and effectively.

The second stage, referred to as the latency-optimisation phase, focuses on identifying the optimal resources for each task category, prioritising the minimisation of delay in task completion. This phase utilises a comprehensive analysis of available resources to allocate them to tasks in ascending order of latency, ensuring that tasks with the most stringent timing requirements are addressed promptly. This is particularly crucial in HIoT environments where delays can lead to performance degradation or system failures. By aligning tasks with the fastest available resources, the PLATOS strategy aims to minimise latency and improve the responsiveness of the system.

\begin{figure}[t!]
    \centering
    \includegraphics[width=100mm]{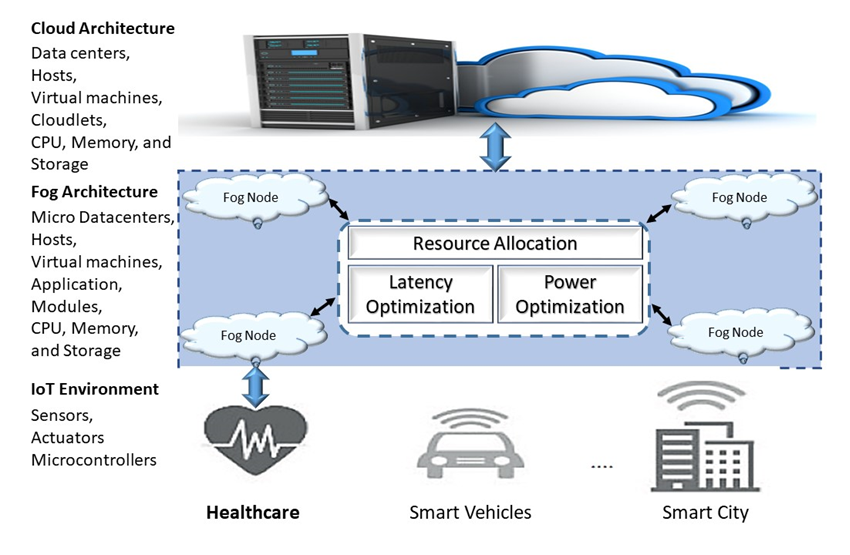}\vspace{-6mm}
    \caption{Proposed PLATOS strategy} \label{Fig: Proposed PLATOS strategy}\vspace{-4mm}
\end{figure}

The third stage termed the power-optimisation phase, involves selecting fog computing resources for each task category based on their power consumption. Resources are sorted in ascending order of power usage, enabling the allocation of energy-efficient resources that align with the power consumption goals of the system. This phase is critical in managing the power demands of HIoT environments, which often involve a multitude of devices and sensors. By optimising power consumption, the PLATOS strategy not only reduces operational costs but also extends the lifespan of the devices involved.

The final stage, the decision-making phase, is responsible for allocating high-performance fog resources to high-priority HIoT tasks. This phase integrates the insights from the previous two optimisation stages to make informed decisions on resource allocation. High-priority tasks, which are identified based on their criticality and urgency, are matched with the most suitable resources to ensure swift and efficient execution. Meanwhile, the remaining tasks are scheduled based on a mapped list of resources from the latency and power optimisation phases. This comprehensive approach ensures that all tasks are executed efficiently while balancing performance, power, and latency considerations.

Algorithm \ref{Alg: Pseudocode of the proposed PLATOS Strategy} demonstrates the proposed PLATOS Strategy.

In the IoT environment, the optimisation of resource allocation strategy is a significant step to achieve energy efficiency, reduce latency, and improve overall system performance. The resource allocation strategy determines how the available resources are allocated to different HIoT tasks to achieve optimal performance with respect to power and latency optimisation. By implementing the PLATOS strategy, it is possible to create a dynamic and responsive IoT ecosystem capable of adapting to varying task demands and resource availabilities.\vspace{-2mm}

\begin{algorithm*}[t!]
{\footnotesize
\textbf{Step 1:} Task Collection\\
HIoT tasks collection: $T = \{t_1, t_2, \dots, t_n\}$\\
HIoT tasks categorisation: $C(T) = \{\{Pt_1, Pt_2, \dots, Pt_n\}+\{St_1, St_2, \dots, St_n\}+\{Comt_1, Comt_2, \dots, Comt_n\}\}$\\
\textbf{Step 2:} Latency Optimisation \\
Latency-aware Resources in ascending order: $L(R) = \{LR_1 < LR_2 < \dots < LR_n\}$\\
\textbf{Step 3:} Power Optimisation \\
Power-aware Resources in ascending order: $P(R) = \{PR_1 < PR_2 < \dots < PR_n\}$\\
\textbf{Step 4:} Resource Mapping \& Decision Making \\
Resource Mapping \& Decision Making: $M(R) = \{\{LR_1 < LR_ 2 < \dots < LR_n\} \cap \{PR_1 < PR_2 < \dots < PR_n\}\}$\\
\textbf{Step 5:} Task Execution \\
Task execution: $TE(M(R)) = \{te_1, te_2, \dots, te_n\}$\\
Status update to fog servers: $SU(TE(M(R))) = \{su_1, su_2, \dots, su_n\}$\\
\textbf{Step 6:} Result Generation\\
Results from fog nodes: $R(SU(TE(M(R)))) = \{r_1, r_2, \dots, r_n\}$\\}
\caption{PLATOS Strategy}\label{Alg: Pseudocode of the proposed PLATOS Strategy}\vspace{-3mm}
\end{algorithm*}

\section{Evaluation Methods}\label{Sec: Evaluation Methods}\vspace{-2mm}

The evaluation of the proposed strategy was conducted using the iFogSim2 simulation environment \cite{mahmud2022ifogsim2}, an enhanced version of the original iFogSim \cite{gupta2017ifogsim}. This advanced simulation tool is tailored for fog and edge computing environments and offers improved capabilities for handling the complexities inherent in such systems. iFogSim2 extends the functionality of its predecessor by incorporating enhanced mobility and clustering features, which are essential for modelling real-world scenarios where IoT devices and edge nodes operate under dynamic conditions.

The primary objective of utilising iFogSim2 in this study was to assess the performance of the proposed strategy under different configurations and operational conditions. The simulator is capable of managing service migration for multiple IoT device mobility models and facilitates the creation of distributed clusters among edge/fog nodes across various hierarchical layers. This orchestration capability allows for a more accurate and flexible simulation of complex environments. The modular design of iFogSim2 enables the selective use of components, such as mobility and clustering, either independently or in combination, to simulate more intricate scenarios.

A key aspect of iFogSim2 is its inclusion of various case studies and test scripts, which enhance its usability and allow researchers to build new rules and scenarios more efficiently. In this study, we leveraged these features to simulate the Cardiovascular Health Monitoring (CHM) application, a widely used application for Electrocardiogram (ECG) monitoring to diagnose heart diseases. The CHM application involves a loop of smart sensors based on Healthcare IoT, which sense and transmit ECG signals to a centralised analysis component. This component then analyses the heart condition based on the received ECG signals. Such applications are designed to execute multiple tasks simultaneously, including filtering ECG data, extracting ECG features, and generating real-time emergency alarms. They also support the long-term collection and analysis of patient data, which is crucial for future health predictions.

The simulations were conducted on a PC with the following configuration: an Intel Core i5 processor running at 2.50 GHz, 16 GB of RAM, and a Windows 10 64-bit operating system. Tab.~\ref{tab:simulation_config} provides a detailed overview of the system configurations used in the simulations.

\begin{table}[b!]
    \centering
    \caption{Simulation System Configuration}\label{tab:simulation_config}
    \begin{tabular}{|l|l|} \hline
        \textbf{Component} & \textbf{Specification} \\  \hline
        Processor & Intel Core i5 2.50 GHz \\ 
        RAM & 16 GB \\
        Operating System & Windows 10 64-bit \\
        Simulation Tool & iFogSim2 \\
        Application & Cardiovascular Health Monitoring (CHM) \\   \hline
    \end{tabular}
\end{table}
To measure the effectiveness of the proposed strategy, we focused on two critical performance parameters: energy consumption and latency. These metrics are crucial for evaluating the efficiency and responsiveness of IoT applications in healthcare settings, where battery-powered devices and real-time data analysis are common.\vspace{-2mm}

\subsection{Energy Consumption}\vspace{-2mm}

Energy consumption is a vital concern in the context of resource allocation for IoT data in fog environments. This metric reflects the amount of energy used by IoT devices and fog nodes during data collection, processing, and transmission. Efficient energy management strategies are crucial to minimise energy consumption and extend the battery life of IoT devices, thereby ensuring sustainable and uninterrupted operation.

The energy consumption in resource allocation for IoT data in fog environments can be calculated using the formula: $E=P\times t$, where $E$ is the energy consumption in Joules (J), $P$ is the power consumption rate in Watts (W), and $t$ is the duration in seconds (s) for which the device is operational. Various factors influence the power consumption rate $P$, including the processing capacity of the device, the volume of data being processed and transmitted, and the communication protocol employed.\vspace{-2mm}

\subsection{Latency}\vspace{-2mm}

Latency refers to the time delay between the generation of data by an IoT device and its processing and analysis by a fog node or cloud server. In the healthcare sector, low latency is crucial as delays in data processing and analysis can lead to serious consequences, such as delayed treatment or misdiagnosis. Ensuring minimal latency is therefore essential for timely and accurate diagnosis and treatment in healthcare IoT applications.

The formula to calculate latency is given by $Latency = T_{end} - T_{start}$,
where Latency is the time delay in seconds (s), $T_{end}$ is the time when the data is processed and analysed by the fog node or cloud server, and $T_{start}$ is the time when the data is generated by the IoT device. Factors affecting latency include the distance between the IoT device and the fog node or cloud server, the processing capacity of the fog node or cloud server, and the communication protocol used.

In conclusion, the proposed PLATOS strategy demonstrates significant improvements in both energy consumption and latency compared to the FTDM strategy. These enhancements are crucial for ensuring efficient and effective operation of healthcare IoT systems, ultimately leading to better patient outcomes and more sustainable healthcare solutions.\vspace{-2mm}

\section{Experiments, Results, and Discussion}\label{Sec: Experiments, Results, and Discussion}\vspace{-2mm}

The iFogSim simulator is modified and used to support the simulation of the proposed PLATOS strategy. The evaluation of the strategy is based on key performance metrics such as energy consumption and latency, which are crucial for the effectiveness of Healthcare IoT tasks. Ensuring minimum latency and reduced energy consumption is paramount in healthcare IoT applications, as it directly influences the quality of patient care and the overall efficiency of healthcare services. The simulation results of the proposed PLATOS strategy are compared with the existing FTDM strategy \cite{saeed2021fault}.

\textbf{Energy Consumption:} Energy consumption is a pivotal concern in healthcare IoT systems due to the typically battery-powered nature of many IoT devices. These devices, which include wearable health monitors and remote sensors, often face constraints related to battery life and energy efficiency. Given that healthcare IoT systems require continuous operation for real-time monitoring and data transmission, efficient energy management becomes crucial.

In healthcare IoT environments, energy consumption directly impacts the operational efficiency and sustainability of IoT devices. High energy consumption can lead to frequent battery replacements or recharges, which not only increases maintenance costs but also interrupts the continuous monitoring of patients. Efficient energy management extends the operational life of these devices, reduces operational costs, and enhances patient and healthcare provider convenience.

\begin{figure}[t!]
\centering
\includegraphics[width=75mm]{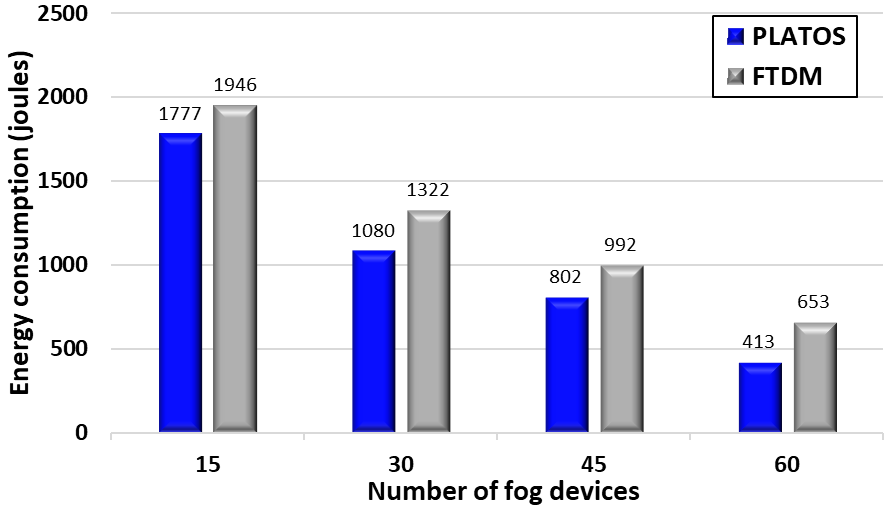} \ \ \includegraphics[width=75mm]{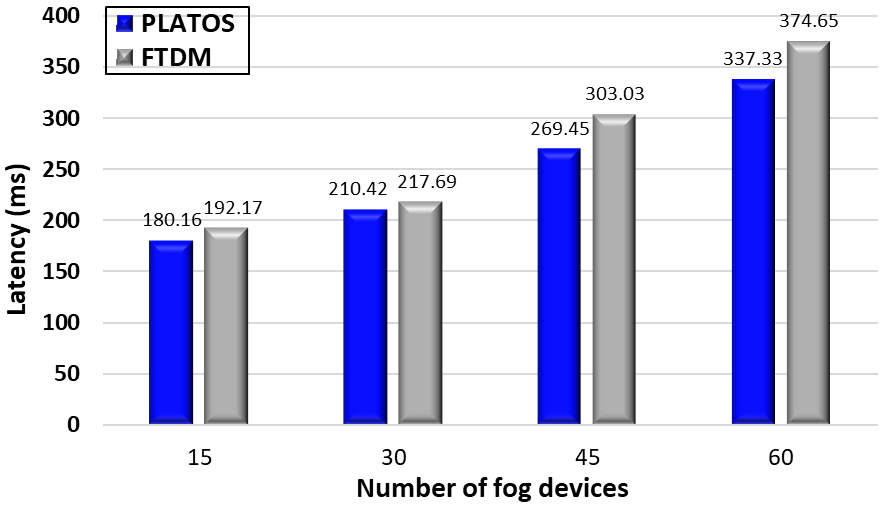} \\[-5mm]
\parbox[t]{75mm}{ \caption{Energy consumption of the proposed and existing strategies} \label{Fig: Energy consumption of the proposed and existing strategies}} \ \ 
\parbox[t]{75mm}{ \caption{Latency of the proposed and existing strategies} \label{Fig: Latency of the proposed and existing strategies}}\vspace{-4mm}
\end{figure}

The proposed PLATOS strategy has been evaluated for its effectiveness in reducing energy consumption compared to the existing FTDM strategy. Fig.~\ref{Fig: Energy consumption of the proposed and existing strategies} displays a comparative analysis of the energy consumption between these strategies. The results indicate that the PLATOS strategy significantly outperforms FTDM, achieving an 18.72\% reduction in energy consumption.

This substantial improvement can be attributed to the PLATOS strategy's efficient task scheduling and resource allocation mechanisms, which are optimised for minimising energy usage. By strategically assigning tasks to nodes based on their energy consumption profiles, the PLATOS strategy ensures that energy is used more efficiently across the network. This results in lower overall energy consumption, as the strategy avoids excessive power drain on any single device and balances the load more effectively.

Moreover, the reduced energy consumption not only contributes to longer device lifetimes but also enhances the overall sustainability of healthcare IoT operations. Lower energy usage translates into fewer interruptions in service due to battery depletion, which is critical for applications that require continuous and reliable monitoring, such as cardiovascular health monitoring.

The effectiveness of the PLATOS strategy in minimising energy consumption underscores its potential to improve the efficiency and sustainability of healthcare IoT systems. By integrating advanced scheduling algorithms that consider energy efficiency as a key parameter, the strategy addresses one of the major challenges in IoT operationsâmanaging energy resources effectively while maintaining high performance and reliability.

\textbf{Latency:} Latency is another vital parameter in healthcare IoT, as the timely transmission of patient data is critical for accurate and prompt decision-making. High latency can lead to delays in diagnosis and treatment, which may have severe implications, including misdiagnosis or inadequate care. In scenarios such as remote patient monitoring, where healthcare providers rely on real-time data to make informed decisions, minimising latency is essential. Efficient latency management ensures that healthcare providers receive the necessary data with minimal delay, thereby facilitating timely interventions and improving patient outcomes. 
Fig.~\ref{Fig: Latency of the proposed and existing strategies} presents the comparative analysis of latency between the PLATOS and FTDM strategies. 

The results indicate that the PLATOS strategy significantly reduces latency by 8.65\%, thus enhancing the responsiveness of healthcare systems and supporting critical decision-making processes.

The proposed PLATOS strategy represents a significant advancement in the scheduling and management of healthcare IoT data tasks. By categorising tasks based on healthcare fields and considering the power consumption of fog and IoT devices, the strategy ensures that tasks are executed efficiently and effectively. The advent of IoT technology is revolutionising the healthcare sector, offering numerous benefits for patients and healthcare professionals alike. Through the deployment of IoT devices such as wearables, sensors, and remote monitoring systems, healthcare providers can collect and analyse data in real-time, facilitating proactive and personalised healthcare services. This technological evolution not only improves patient outcomes by enabling early detection of health issues but also helps in preventing hospital readmissions and reducing healthcare costs. 

Furthermore, IoT-enabled healthcare devices empower patients to manage chronic conditions and adopt healthier lifestyles. Wearable devices that track physical activity, sleep patterns, and vital signs encourage individuals to maintain regular exercise routines, monitor their health, and make informed decisions about their well-being. IoT technology also enhances the operational efficiency of healthcare systems by automating tasks, optimising resource allocation, and reducing waiting times, thereby improving the overall quality of patient care. The implementation of IoT-enabled healthcare devices facilitates remote consultations, telemedicine, and virtual care, expanding access to healthcare services for individuals in remote or underserved areas. The transformative potential of IoT technology in the healthcare industry is immense, providing significant benefits in terms of improved patient outcomes, enhanced disease management, streamlined healthcare operations, and increased access to healthcare services.

The simulation-based evaluation of the proposed PLATOS strategy demonstrates its efficacy in reducing energy consumption and latency by 18.72\% and 8.65\% respectively, compared to the existing FTDM strategy. This improvement is attributed to the strategic optimisation of fog and IoT device power by allocating IoT tasks to nodes that consume the least power. Additionally, the PLATOS strategy is designed to handle tasks as per their resource requirements by assigning them to the most suitable resource, thereby minimising latency and enhancing overall system performance.

\textbf{Device Efficiency Index (DEI):} The DEI is a metric designed to evaluate the efficiency of fog devices in terms of both energy consumption and latency. The DEI is calculated by the following formula:\vspace{-2mm}
\begin{equation}
\text{DEI} = \frac{\text{Number of Fog Devices}}{\text{Energy Consumption} \times \text{Latency} + \epsilon},\vspace{-3mm}
\end{equation}
where:\vspace{-2mm}
\begin{itemize}\itemsep=-2pt
    \item \textbf{Number of fog devices} represents the number of fog nodes available in the network.
    \item \textbf{Energy Consumption} is the total energy consumed by the fog nodes.
    \item \textbf{Latency} is the time delay associated with processing and transmitting tasks within the fog network.
    \item \(\epsilon\) is a small constant added to avoid division by zero, typically set to \(10^{-6}\).
\end{itemize}

\begin{figure}[t!]
    \centering
    \includegraphics[width=75mm]{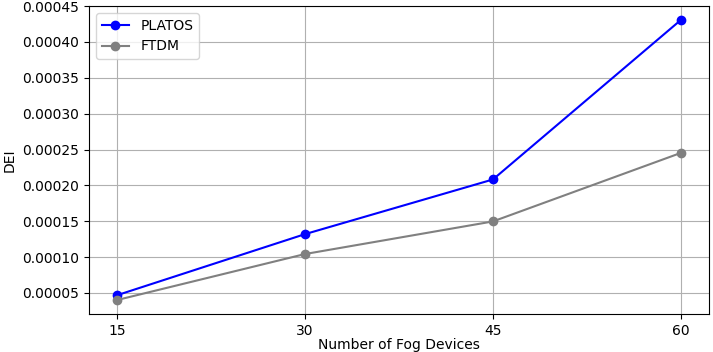} \ \  \includegraphics[width=72mm]{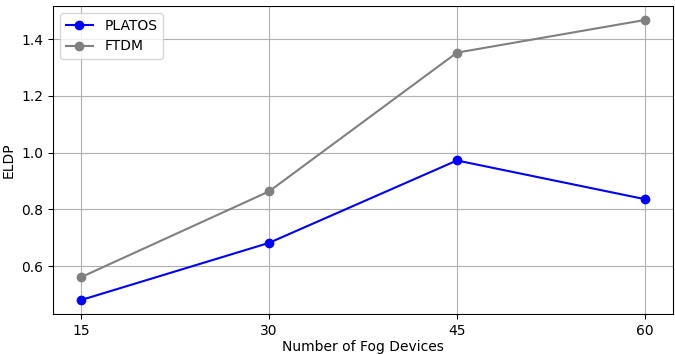} \\[-5mm]
\parbox[t]{75mm}{ \caption{DEI vs. Number of Fog Devices} \label{Fig: DEI}} \ \ \
\parbox[t]{72mm}{ \caption{ELDP vs Fog Devices} \label{Fig: ELDP}}\vspace{-4mm}
\end{figure}

Fig.~\ref{Fig: DEI} shows the DEI for both the proposed system and the FTDM across different numbers of fog devices.

The key observations are as follows:
\begin{itemize}\itemsep=-2pt
    \item \textbf{Initial Stages (15 and 30 Devices):} The DEI values are relatively low, indicating that with fewer fog devices, both the proposed and benchmark systems exhibit less efficiency in reducing energy consumption and latency. This might be due to the underutilisation of available resources.
    \item \textbf{Mid to High Stages (45 and 60 Devices):} As the number of fog devices increases, the DEI values rise significantly for the proposed system compared to the benchmark. This suggests that the proposed system becomes more efficient at utilising fog resources to optimise both energy consumption and latency as more devices are deployed.
\end{itemize}

The DEI plot demonstrates that the proposed system achieves better efficiency than the benchmark system, particularly as the network scales with more fog devices. This indicates the effectiveness of the proposed task scheduling approach in managing energy and latency, making it more suitable for large-scale IoT deployments.

\textbf{Energy-Latency-Device Product (ELDP):} The ELDP is proposed as a comprehensive metric to evaluate the efficiency of fog computing systems. It combines three crucial factors: energy consumption, latency, and the number of fog devices. The ELDP is defined as the product of the total energy consumption, the average latency, and the number of fog devices:
\begin{equation}
\text{ELDP} = \text{Energy} \times \text{Latency} \times \text{Number of Fog Devices}
\end{equation}

This metric provides an overall measure of the system's efficiency, where lower ELDP values indicate a more efficient system.

The ELDP values for both the proposed system and the benchmark system were computed and plotted against the number of fog devices. As illustrated in Fig.~\ref{Fig: ELDP}, the following observations can be made:
\begin{itemize}\itemsep=-2pt
    \item \textbf{Scalability}: The ELDP generally increases with the number of fog devices, reflecting the cumulative impact of adding more devices to the system. This is expected, as more devices typically introduce additional energy consumption and latency. However, the proposed system shows a more controlled increase in ELDP, particularly when the number of devices exceeds 45. This suggests that the proposed system is more scalable and efficient as it adds more fog devices.
    \item \textbf{Comparison with Benchmark}: The proposed system consistently exhibits a lower ELDP compared to the benchmark system. This indicates that the proposed system is more efficient in managing energy consumption and latency, especially as the number of fog devices increases. The benchmark system shows a linear increase in ELDP, highlighting its lower efficiency in handling the increased computational load.
    \item \textbf{Optimal Device Count}: Notably, the proposed system's ELDP increases up to 45 devices and then slightly decreases as the number of devices reaches 60. This could indicate an optimal number of fog devices for the proposed system, where the balance between energy consumption and latency is most favourable. In contrast, the benchmark system continues to show an upward trend, suggesting less efficiency beyond a certain point.
    \item \textbf{Implications}: The ELDP metric emphasises the importance of balancing energy consumption and latency in fog computing environments. The proposed system's ability to maintain a lower ELDP across different numbers of fog devices highlights its potential for practical deployment in scenarios requiring efficient resource management.
\end{itemize}

The ELDP metric provides a holistic view of the performance of fog computing systems, particularly in terms of energy and latency management across varying numbers of fog devices. The analysis shows that the proposed system is superior in maintaining low energy consumption and latency as the system scales, making it a more viable solution for large-scale deployments.\vspace{-2mm}
    
\section{Conclusion}\label{Sec: Conclusion}\vspace{-2mm}

In this study, a PLATOS approach is suggested for Healthcare IoT tasks. The PLATOS plan was created with a four-stage implementation in mind. It divided the tasks related to the HIoT into three different categories in the first step, known as the task-orientation phase. Priority-oriented activities, storage-oriented tasks, and computationally-oriented tasks are the three categories mentioned. The second step, referred to as the latency-optimisation phase, is where the best resources are found for each category of tasks in ascending order according to the shortest delay when the tasks are performed. The third phase, referred to as the power-optimisation phase, involves selecting the optimal computing resources for each category of activity, ranking them according to least power usage. The decision-making step, which comes after the third and final phase, is in charge of distributing high-performance fog resources to the high-priority HIoT tasks. The latency and power-optimisation stages provide a mapped list of resources, which is used to schedule the remaining HIoT jobs.

Simulations were run in iFogSim2, and the results show that the proposed PLATOS method reduces energy consumption and latency in comparison to the present state-of-the-art strategy by 18.72\% and 8.65\%, respectively. Furthermore, the proposed system demonstrated improved efficiency through the DEI, indicating better utilisation of fog resources with increased numbers of fog devices. Additionally, the ELDP provided a comprehensive measure of the system's performance, showing that the proposed approach is more scalable and efficient, particularly as the system scales with more fog devices.

In the future, the proposed strategy will be further strengthened by implementing a fault-tolerant mechanism. Additionally, future work will focus on integrating machine learning techniques to enhance task classification and resource allocation dynamically. The scalability of the PLATOS strategy will also be tested in large-scale real-world HIoT deployments to validate its effectiveness beyond simulation environments.\vspace{-4mm}


\footnotesize
\noindent
\parbox[t]{75mm}{Mohammed Alaa Ala'anzy,\\
Department of Computer Science, SDU University,\\
040900 Kaskelen, Kazakhstan\\
Email: {\tt m.alanzy@ieee.org},\\
} \ \
\parbox[t]{75mm}{Zulfiqar Ahmad,\\
Department of Computer Science and Information Technology, Hazara University, \\
21300 Mansehra, Pakistan\\
Email: {\tt zulfiqarahmad@hu.edu.pk},} \\[3mm]
\parbox[t]{75mm}{Zhanar Mukash,\\
Department of Computer Science, SDU University,\\
040900 Kaskelen, Kazakhstan \\
Email: {\tt zhanar.mukash@sdu.edu.kz}}

\vspace{5mm}
Received  22.01.2025, $\> \> \> $ Revised 11.02.2025, $\> \> \> $   Accepted 12.02.2025,  $\> \> \> $    Available online 30.09.2025.

\end{document}